\documentclass[a4paper]{revtex4}
\usepackage{graphicx}
\usepackage{fancyhdr}
\usepackage{amsmath}
\usepackage{color}

\begin{document}

\title{Standard Model criticality and Higgs inflation}

%

\author{Yuta Hamada}
\affiliation{Department of Physics, Kyoto University, Kyoto 606-8502, Japan}

\begin{abstract}
It is known that the Higgs potential becomes flat around the Planck scale under the assumption that the Standard Model is valid up to very high scale.
Taking this into account, we revisit the Higgs inflation scenario and find that the various types of inflation occur depending on the value of the top mass.
We also discuss the implication from non-supersymmetric string theory.   
\end{abstract}

\maketitle

\thispagestyle{fancy}


\section{Introduction}
The observed Higgs mass leads to the fact that the Higgs self coupling and its beta function become zero at string/Planck scale depending on the value of the top mass~\cite{flat potential,Hamada:2012bp}.
Further, bare Higgs mass also vanish around this scale~\cite{flat potential,Hamada:2012bp}.
These facts indicate that the Higgs potential is very flat, which is good for the inflation by the Higgs boson~\cite{Bezrukov:2007ep,Hamada:2014iga,critical,Haba:2014zda}.
By introducing the non-minimal coupling between the Higgs field and the Ricci scalar $\xi |H|^2 \mathcal{R}$~\cite{Salopek:1988qh}, the Higgs potential becomes sufficiently flat to realize the successful inflation. For the critical case, only $\xi= O(10)$ is needed~\cite{Hamada:2014iga,critical,Haba:2014zda}, while $\xi=O(10^4)$ is needed for non-critical case.
Furthermore, if the multiple point criticality principle~\cite{Froggatt:1995rt} is satisfied, the Higgs field becomes the seeds of the topological inflation~\cite{Hamada:2014raa}.
This paper is based on our works~\cite{Hamada:2012bp,Hamada:2014raa,Hamada:2014iga,Hamada:2014xka,Hamada:2014ofa,Hamada:2015ria}.



\section{Higgs potential and inflation}
\begin{figure}[ht]
\centering
\includegraphics[width=80mm]{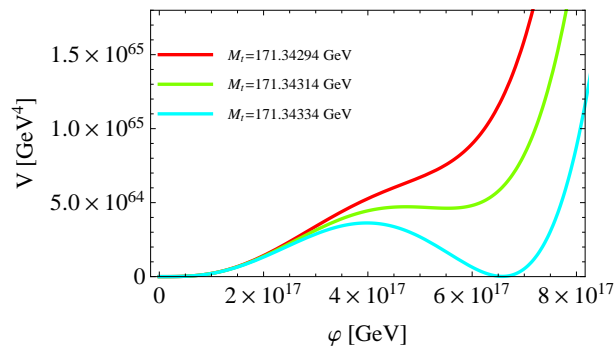}
\caption{The Higgs potential as a function of the Higgs field value $\varphi$ near the SM criticality. Here we take $M_H=125.9$GeV and $\alpha_s=0.1185$ and use the two loop renormalization group equations.} \label{SM potential}
\end{figure}

At the scale much larger than the electroweak scale, we can approximate the Standard Model(SM) Higgs potential as 
\begin{equation}
V\sim {\lambda(\varphi)\over4} \varphi^4,
\end{equation}
with $\varphi$ being the physical Higgs field.
In Fig.~\ref{SM potential}, we plot the Higgs potential as a function of $h$ with various values of top mass $M_t$.
See e.g. Ref.~\cite{Kawabataa:2014osa} for the current uncertainty of the top mass.
We can see that our vacuum is stable for lighter $M_t$ and is unstable for heavier $M_t$. 
Furthermore, if $M_t$ is tuned, there arises the saddle point where the first and the second derivatives of the potential is zero. 
In the following, we introduce the Higgs inflation in these three cases.

\subsection{Usual Higgs inflation and the saddle point inflation}
For the small value of $M_t$, the Higgs potential is stable up to the inflation scale and, we can realize successful inflation. 
By adding the non-minimal coupling $\xi$, the action becomes
\begin{align}
S=\int \sqrt{-g}\, d^4x
\left(
-M_P^2 \mathcal{R}+\xi \varphi^2 \mathcal{R}+{1\over2}(\partial_\mu \varphi)^2-
{\lambda\over4}\varphi^4
\right).
\end{align}
Here we focus on the Higgs and gravity sectors and neglect other sectors for simplicity.
After the redefinition of the metric, 
\begin{align}
&
g_{E\mu\nu}=\Omega^2 g_{\mu\nu},
&
\Omega^2=1+{\xi\varphi^2\over M_P^2},
\end{align}
we obtain the following Einstein frame action,
\begin{align}
S=\int \sqrt{-g_E}\, d^4x
\left(
-M_P^2 \mathcal{R}_E+{1\over2}(\partial_\mu \chi)^2-
{\lambda\over4}{\varphi^4\over(1+\xi \varphi^2/M_P^2)^2}
\right),
\end{align}
where $\chi$ is the canonical field in the Einstein frame, and the relation between $\varphi$ and $\chi$ is
\begin{align}
{d\chi\over d\varphi}=\sqrt{\left(\Omega^2+{6\xi^2\varphi^2\over M_P^2}\right){1\over\Omega^4}}.
\end{align}
The point is that the potential is divided by the factor $\Omega^4$.
This fact leads the flat potential because $\varphi^4/(1+\xi\varphi^2/M_P^2)^2$ becomes constant at the high scale.
For the value of $\lambda=O(0.1)$, $\xi$ is needed to be $O(10^4)$ to fit the COBE normalization.

If the $M_t$ is tunes so that the Higgs potential becomes flat, the required amount of the non-minimal coupling $\xi$ is significantly reduced since the potential is suppressed due to the smallness of $\lambda$.
This corresponds to the green line in Fig.~\ref{SM potential}.
We only need small $\xi=O(10)$ to realize the inflation~\cite{Hamada:2014iga,critical}.
%





\subsection{topological Higgs inflation}
\begin{figure}[ht]
\centering
\includegraphics[width=80mm]{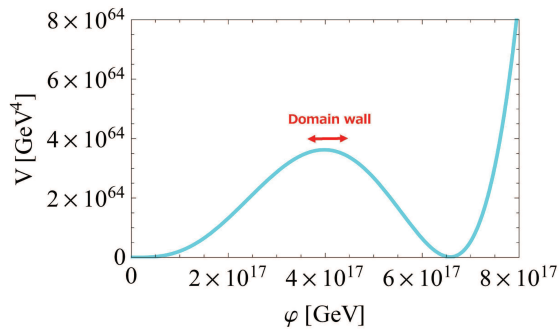}
\caption{
The schematic picture of the formation of the domain wall in the Higgs potential.
} \label{domain wall}
\end{figure}

Let us focus on the case where there are two vacua as in blue line in Fig.~\ref{SM potential}, which is predicted by Froggatt and Nielsen~\cite{Froggatt:1995rt} about twenty years before the Higgs discovery. 
See also Refs.~\cite{MEP,Hamada:2014ofa} for another explanation of the SM criticality, and for Refs.~\cite{extension,Hamada:2014xka} for the criticality in the minimal extensions of the SM.
Since the potential has two vacua, the domain wall connecting the our vacuum and the Planck scale one would be formed in general initial condition, see Fig.~\ref{domain wall}.

In this case, the topological inflation driven by the Higgs field is possible if the thickness of the domain wall is larger than the Hubble horizon case~\cite{Hamada:2014raa}. See also Ref.~\cite{Kitajima:2015nla} for the production of the Higgs domain wall after the inflation.
The thickness of the domain $\delta_\text{DW}$ wall can be estimated as follows.
Let us start from the action,
\begin{align}
S=\int d^4x \left(
                 (\partial \varphi)^2- V(\varphi)
                 \right) .
\end{align}
If we consider the the domain wall configuration which we denote $\varphi_\text{DW}$,
the terms in the action should be balanced.
\begin{align}
(\partial \varphi_\text{DW})^2\simeq V(\varphi_\text{DW}).
\end{align}
Naively,  $\partial \varphi_\text{DW}\simeq \varphi_\text{DW}/\delta_\text{DW}$,
and we get
\begin{align}
\delta_\text{DW}\simeq {V\over \varphi_\text{DW}^2}\simeq V''(\varphi_\text{DW}).
\end{align}
Practically, we evaluate $V''(\varphi_\text{DW})$ at the local maximum of the Higgs potential.

The condition for the topological Higgs inflation is satisfied by adding small non-minimal coupling $\xi=O(0.1-10)$ or the right handed neutrino~\cite{Hamada:2014raa}.




\section{Higgs potential beyond the Planck scale}
We comment on the implication from string theory.
Higgs potential beyond the Planck scale can not be computed by using field theory.
In order to discuss this, we consider the non-supersymmetric string theory~\cite{SO(16) model}. This is because that the tachyon-free non-supersymmetric vacua is generic in string theory~\cite{fermionic construction} and that
the recent LHC result may indicate that the study of the phenomenology based on non-supersymmetric string~\cite{nonSUSY string, Hamada:2015ria} become important.
The flat potential around the string scale suggests that the Higgs comes from the tree level massless state of string.
The Higgs field generally mixes other moduli above the Planck scale, and 
we can argue that at least the one of the direction, radion direction, has runaway vacuum if the ten dimensional cosmological constant is positive using the non-supersymmetric string model.

Taking into account this, we conclude that the topological Higgs inflation is generically realized in string theory~\cite{Hamada:2015ria}. 


\section{Summary}
We have considered the Higgs inflation in light of the discovery of the Higgs boson.
Taking into account the small $\lambda$ at high scale, we can realize the Higgs inflation with small non-minimal coupling.
We also have discussed the topological inflation by the Higgs field.
These inflations are compatible with non-supersymmetric string theory.
\begin{acknowledgments}
We thank Hikaru Kawai, Kin-ya Oda, Seong Chan Park and Fuminobu Takahashi for useful discussion and fruitful collaborations.
\end{acknowledgments}

\bigskip 

\end{document}